\documentclass[superscriptaddress,showpacs,amssymb,10pt,aps,prd,longbibliography,reprint,footinbib]{revtex4-1}

\usepackage{graphicx,epsfig,amssymb,times} 
\usepackage{amsmath,amsfonts}
\usepackage{bm}
\usepackage{epstopdf}
\usepackage[linktocpage,colorlinks]{hyperref}
\usepackage[caption=false]{subfig}
\usepackage[usenames]{color}     

\definecolor{coolblack}{rgb}{0.0, 0.18, 0.39}
\definecolor{darkred}{rgb}{0.5,0,0}
\definecolor{darkgreen}{rgb}{0,0.5,0}
\definecolor{darkblue}{rgb}{0,0,0.5}
\definecolor{lapislazuli}{rgb}{0.15, 0.38, 0.61}
\definecolor{venetianred}{rgb}{0.78, 0.03, 0.08}
\definecolor{bleudefrance}{rgb}{0.19, 0.55, 0.91}
\definecolor{dogwoodrose}{rgb}{0.84, 0.09, 0.41}
\hypersetup{colorlinks=true, citecolor=darkblue, linkcolor=darkblue, 
urlcolor = darkblue}

\def\be{\begin{equation}}
\def\ee{\end{equation}}

\renewcommand{\vec}[1]{\boldsymbol{#1}}

\newcommand{\bea}{\begin{eqnarray}}
\newcommand{\eea}{\end{eqnarray}}
\newcommand{\ben}{\begin{enumerate}}
\newcommand{\een}{\end{enumerate}}
\newcommand{\bi}{\begin{itemize}}
\newcommand{\ei}{\end{itemize}}

\renewcommand{\d}{\partial}

\def\ga{\mathrel{\raise.3ex\hbox{$>$\kern-.75em\lower1ex\hbox{$\sim$}}}}
\def\la{\mathrel{\raise.3ex\hbox{$<$\kern-.75em\lower1ex\hbox{$\sim$}}}}

\def\l{\left}
\def\r{\right}
\def\be{\begin{equation}}
\def\ee{\end{equation}}

\renewcommand{\vec}[1]{\boldsymbol{#1}}

\renewcommand{\d}{\rm{d}}

\def\I_M{{I_{\scriptscriptstyle M\times M}}}

\def\be{\begin{equation}}
\def\ee{\end{equation}}
\def\bea{\begin{eqnarray}}
\def\eea{\end{eqnarray}}
\newcommand{\beq}{\begin{eqnarray}}
\newcommand{\eeq}{\end{eqnarray}}

\newcommand{\beqal}{\begin{eqnarray}\label}
\newcommand{\beqa}{\begin{eqnarray}}
\newcommand{\eeqa}{\end{eqnarray}}

\newcommand {\non}{\nonumber\\}
\newcommand {\f}{\frac} 
\newcommand {\swf}{\mathtt{S}} 
\newcommand {\rtf}{\mathtt{R}} 
\newcommand {\slmw}{\mathfrak{s}lm\omega}
\newcommand {\lmw}{lm\omega}
\newcommand {\sw}{\mathfrak{s}}
\newcommand {\aw}{\mathfrak{c}}

\newcommand{\omtil}{\widetilde{\omega}}

\begin{document}
\title{\large Absorption of electromagnetic plane waves by rotating black holes}

\author{Luiz C. S. Leite}
\email{luizcsleite@ufpa.br}
\affiliation{Faculdade de F\'{\i}sica, Universidade 
Federal do Par\'a, 66075-110, Bel\'em, Par\'a, Brazil.}
\author{Sam Dolan}
\email{s.dolan@sheffield.ac.uk}
\affiliation{Consortium for Fundamental Physics, School of 
Mathematics and Statistics, University of Sheffield, Hicks Building,
Hounsfield Road, Sheffield S3 7RH, United Kingdom}
\author{Lu\'is C. B. Crispino}
\email{crispino@ufpa.br}
\affiliation{Faculdade de F\'{\i}sica, Universidade 
Federal do Par\'a, 66075-110, Bel\'em, Par\'a, Brazil.}

\date{\today}


\begin{abstract}
We study the absorption of monochromatic electromagnetic plane 
waves impinging upon a Kerr black hole, in the general case that the direction of incidence is  
not aligned with the black hole spin axis. We present numerical results that are in accord with low- and high-frequency approximations. We find that circularly-polarized waves are distinguished by the black hole spin, with counter-rotating polarizations being more absorbed than co-rotating polarizations. At low frequencies and moderate incidence angles, there exists a narrow parameter window in which superradiant emission in the dipole mode can exceed absorption in the non-superradiant modes, allowing a planar electromagnetic wave to stimulate net emission from a black hole.
\end{abstract}

\pacs{
04.70.-s, 
04.70.Bw, 
11.80.-m, 
04.30.Nk, 
}
\maketitle

\section{Introduction}
The recent announcements from the LIGO and Virgo detectors have confirmed a key prediction of General Relativity~(GR): orbiting companions generate gravitational waves which propagate at the speed of light. The gravitational waves observed, so far, are outcomes of catastrophic 
events apparently involving black holes (BHs) and neutron stars~\cite{abbott2016observation,abbott2016gw151226,scientific2017gw170104,abbott2017gw170608,abbott2017gw170814,abbott2017gw170817}. Gravitational waves from the BH `ringdown' phase probe the strong-field regime of GR for the first time. BHs are fascinating objects which play a central role in modern astrophysics, and which provide a crucible for new tests of physical theory. 

As a BH possesses an event horizon, it can absorb matter, energy and radiation in its vicinity. To understand this process in detail, one may study an idealised scenario: the absorption of a planar wave of a fixed angular frequency $\omega$ and spin $s$ impinging upon a BH of mass $M$ and angular momentum $J$. Absorption is characterised by a single key quantity, the absorption cross section $\sigma_{abs}$, expressed in terms of dimensionless parameters such as $s$ and $M\omega \equiv \pi r_s / \lambda$, where $r_s$ is the Schwarzschild radius of the body, and $\lambda$ is the wavelength.

The foundations of planar-wave scattering theory on BHs spacetimes were set out in Ref.~\cite{Futterman:1988ni}. That work addressed the cases of monochromatic scalar ($s=0$), neutrino ($s=1/2)$, electromagnetic ($s=1$) and gravitational waves ($s=2$) by Kerr BHs, with particular focus on the low-frequency ($M \omega \ll 1$) and high-frequency ($M\omega \gg 1$) regimes, which are amenable to asymptotic methods \cite{Futterman:1988ni}. 

The intermediate regime $M\omega \sim 1$ is typically studied via numerical methods. 
The first numerical investigations were carried out by Sanchez in the 1970s~\cite{Sanchez:1977si,Sanchez:1977vz} for Schwarzschild BHs (see also \cite{Page:1976df, Page:1977um}). Subsequently, various authors have  
computed the scalar~\cite{jung2005absorption,Crispino:2009ki,benone2014absorption,PhysRevD.90.064001,
PhysRevD.92.024012,benone2016superradiance}, 
fermionic~\cite{Unruh:1976fm, Doran:2005vm,dolan2007scattering,cotuaescu2016partial}, 
electromagnetic~\cite{Crispino:2007qw,Crispino:2008zz,Crispino:2009xt}, and 
gravitational~\cite{Dolan:2008kf} absorption cross sections for static BHs solutions. 

\begin{figure*}
\includegraphics[height=7.2cm]{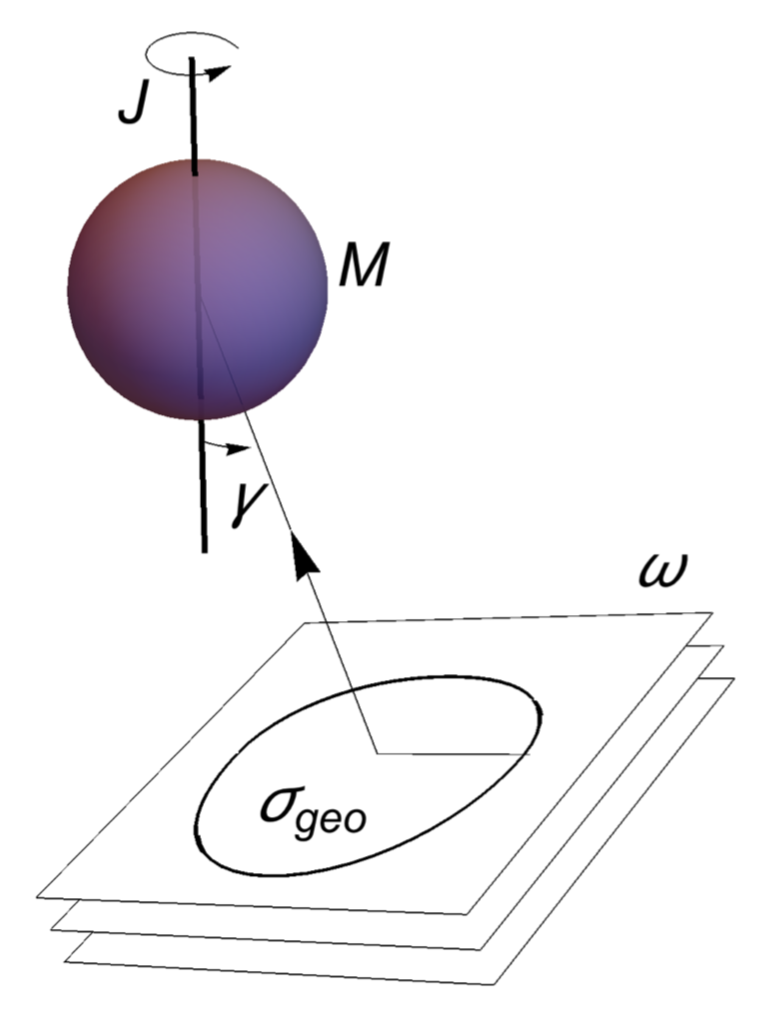} \quad 
\includegraphics[height=7.5cm]{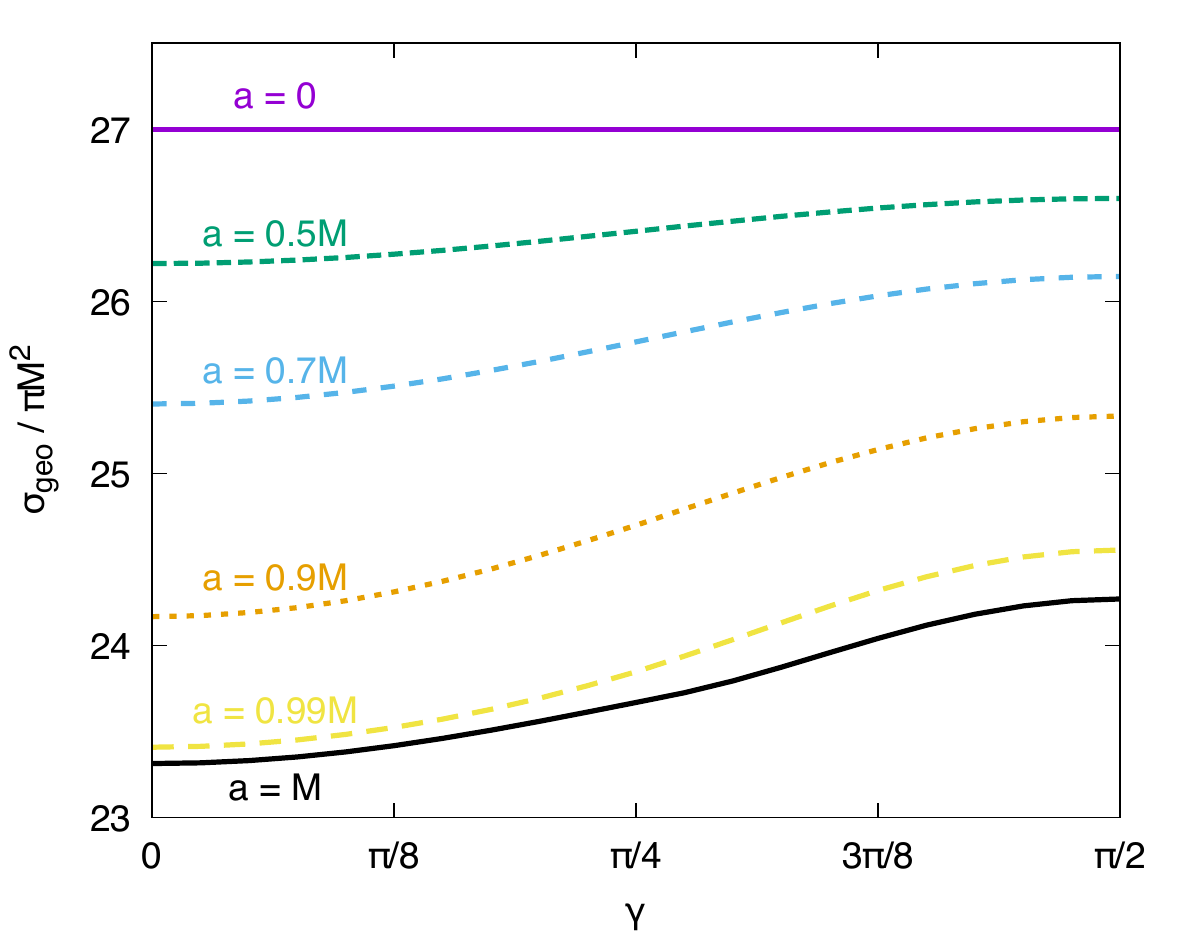}
\caption{\emph{Left:} A monochromatic wave of angular frequency $\omega$ impinges upon a Kerr black hole of mass $M$ and angular momentum $J$ at an angle of $\gamma$ relative to the spin axis. \emph{Right:} The geodesic capture cross section $\sigma_{geo}$ as a function of  the angle of incidence $\gamma$ for various spin rates $a \equiv J/M$.}
\label{fig:sketch}
\end{figure*}

There is strong evidence that astrophysical BHs are not static, but instead possess significant angular momentum. To describe the absorption scenario on the axially-symmetric Kerr spacetime one needs additional dimensionless parameters: the BH spin ratio, $0 \le J / M^2 < 1$; the angle between the incident direction and the symmetry axis, $0 \le \gamma < \pi$; and the polarization state of the wave $\mathcal{P}$ (see Fig.~\ref{fig:sketch}).

Here we seek to fill a gap in the literature, by calculating the absorption cross section for planar electromagnetic waves incident on rotating (Kerr) BHs via numerical methods. This work extends Ref.~\cite{Leite:2017zyb}, which restricted attention to the case of on-axis incidence ($\gamma = 0$), and it complements numerical investigations of the scalar~\cite{glampedakis2001scattering,Caio:2013}, fermionic~\cite{dolan2007scattering}, and gravitational~\cite{Dolan:2008kf} cases for Kerr BHs.

First, let us review the key results in the limiting regimes. At long wavelengths ($M\omega \ll 1$), absorption occurs primarily in the dipole modes of the electromagnetic field ($l=1$). In the Schwarzschild case ($J = 0$), the electromagnetic absorption cross section is simply $\sigma_{abs} \approx \frac{4}{3} \mathcal{A}_S (M \omega)^2$ \cite{Crispino:2007qw}, where $\mathcal{A}_S = 4 \pi (2 M)^2$ is the area of the Schwarzschild event horizon. Thus, the BH does not substantially absorb low-frequency modes of the electromagnetic field, that is, modes whose wavelength is significantly longer than the black hole's horizon radius. In contrast, long-wavelength scalar fields are substantially absorbed, as $\sigma^{s=0}_{abs} \approx \mathcal{A}$ \cite{Das:1996we, Higuchi:2001si}. 

A rotating BH can actually amplify, rather than absorb, low-frequency waves \cite{starobinsky1973aa,Rosa:2016bli}. For a long-wavelength circularly-polarized wave incident along the black hole rotation axis,
\beq
\sigma_{abs} \approx \frac{4}{3} \mathcal{A} \, \omega (\omega - \Omega_H) M^2, \label{eq:lowfreq}
\eeq
where $\Omega_H = J / 2M^2r_+$ is the angular frequency of the event horizon of radius $r_+$, and $\mathcal{A} = 8 \pi M r_+$ is the horizon area \cite{Page:1976df}. The co- and counter-rotating polarizations correspond to the  frequencies $+|\omega|$ and $-|\omega|$, respectively. The absorption cross section is negative for modes in the superradiant regime.

Superradiance may be anticipated from the laws of black hole mechanics \cite{Bardeen:1973gs}. The second law states that the area of the black hole is a non-decreasing function of time, if the weak energy condition holds. The first law may be written in the form $d\mathcal{A} = \left(1 - \Omega_H \frac{dJ}{dM} \right) 8 \pi \kappa^{-1} d M$, where $\kappa > 0$ is the surface gravity of the horizon. Associating $dJ / dM$ with $m / \omega$, the first law implies that $dM \leq 0$ whenever $\omega \omtil < 0$, since $d \mathcal{A} \geq 0$ by the second law.

At short wavelengths ($M\omega \gg 1$), the geometrical-optics approximation is valid, and one may employ ray-tracing methods. The cross section $\sigma_{abs}$ approaches the geodesic capture cross section, $\sigma_{geo}$, defined by the area on the initial wavefront whose boundary is defined by the set of null geodesics which are neither scattered nor absorbed by the hole, but which instead orbit indefinitely \cite{Chandrasekhar:1998, Caio:2013}. The geodesic capture cross section is shown in Fig.~\ref{fig:sketch}.

This paper is arranged as follows. In Sec.~\ref{sec:metric} 
we review the Kerr solution in Boyer-Lindquist coordinates and the basic equations for 
perturbing fields around in the Kerr background. In Sec.~\ref{sec:absorption} we show an expression for the 
absorption cross section obtained via the partial-wave method. In Sec.~\ref{sec:numerical_method} 
we detail the numerical methods used to compute the absorption cross section. 
We present our numerical results in Sec.~\ref{sec:num_results}, and we conclude in Sec.~\ref{sec:remarks} with some final remarks. In this paper we assume natural units such that $G = c = 1$.

\section{Kerr spacetime and Teukolsky equations}\label{sec:metric}
It is well known that in GR, four-dimensional neutral rotating BHs 
are described by the Kerr solution. Kerr 
BHs are characterized by two parameters: mass $M$ and angular momentum $J$, with the latter commonly represented by $a =J/M$, the angular momentum per unit mass. 

The Kerr solution in the Boyer-Lindquist coordinates \{$t,r,\theta,\varphi$\} has the line element 
\bea
{\d} s^2&=& -\f{\Delta}{\Sigma}\l({\d} t-a\sin^2\theta{\d} \varphi \r)^2 + \frac{\Sigma}{\Delta}{\d} r^2 + \Sigma\,{\d}\theta^2 + 
\nonumber\\
&& + \f{\sin^2\theta}{\Sigma}\l[(r^2+a^2){\d}\varphi-a{\d}t\r]^2,\label{eq:linelement}
\eea
with~$\Sigma\equiv r^2+a^2\cos^2\theta$, and $\Delta\equiv r^2-2Mr+a^2$. When the 
condition $a^2\leq M^2$ is satisfied, the Kerr solution corresponds to a BH spacetime. We will restrict attention to the case~$a^2 < M^2$, which corresponds to a rotating BH with two 
distinct horizons: an internal~(Cauchy) horizon located at~$r_{-}=M-\sqrt{M^2-a^2}$ and 
an external~(event) horizon at~$r_{+}=M+\sqrt{M^2-a^2}$.

Using the Newman-Penrose formalism, Teukolsky gathered the description of perturbing fields, in the Kerr BH spacetime, into a single master equation that, in the vacuum, reads \cite{teukolsky1972rotating}
\bea
\left[\frac{(r^{2}+a^{2})^{2}}{\Delta}-a^{2}\sin^{2}\theta\right]\frac{\partial^{2}\Upsilon_\mathfrak{s}}{\partial t^{2}}+\frac{4Mar}{\Delta}\frac{\partial^{2}\Upsilon_\mathfrak{s}}{\partial t\partial\varphi}\non+\left[\frac{a^{2}}{\Delta}-\frac{1}{\sin^{2}\theta}\right]\frac{\partial^{2}\Upsilon_\mathfrak{s}}{\partial\varphi^{2}}
-\Delta^{-\mathfrak{s}}\frac{\partial}{\partial r}\left(\Delta^{\mathfrak{s}+1}\frac{\partial\Upsilon_\mathfrak{s}}{\partial r}\right)\non-\frac{1}{\sin\theta}\frac{\partial}{\partial\theta}\left(\sin\theta\frac{\partial\Upsilon_\mathfrak{s}}{\partial\theta}\right)+(\mathfrak{s}^{2}\cot^{2}\theta-\mathfrak{s})\Upsilon_\mathfrak{s}\nonumber \\
-2\mathfrak{s}\left[\frac{a(r-M)}{\Delta}+\frac{i\cos\theta}{\sin^{2}\theta}\right]\frac{\partial\Upsilon_\mathfrak{s}}{\partial\varphi}\non-2\mathfrak{s}\left[\frac{M(r^{2}-a^{2})}{\Delta}-r-ia\cos\theta\right]\frac{\partial\Upsilon_\mathfrak{s}}{\partial t}=0,\label{eq:master_eq}
\eea
where $\mathfrak{s}$ is the spin-weight of the field, and we have ${\mathfrak{s}}=0,\,\pm1/2,\,\pm1,\,\pm2$, for scalar, neutrino, electromagnetic and gravitational perturbations, respectively. 

Our focus here is on electromagnetic waves; hence we choose $\mathfrak{s}=-1$, noting that $\Upsilon_{-1} \equiv \phi_2 / (r-ia\cos\theta)^2$, where $\phi_2 \equiv F_{\mu \nu} \overline{m}^\mu n^\nu$ is a Maxwell scalar, $F_{\mu \nu}$ is the Faraday tensor, and $\overline{m}^\mu$ and $n^\mu$ are legs of Kinnersley's null tetrad \cite{teukolsky1972rotating}.

Following Teukolsky's route \cite{teukolsky1972rotating, teukolsky1973perturbations, press1973perturbations, teukolsky1974perturbations}, one separates variables in Eq~\eqref{eq:master_eq} using the following ansatz
\be
\Upsilon_{-1\lmw}(t,\,r,\,\theta,\,\varphi)=\rtf_{-1\lmw}(r)\swf_{-1\lmw}(\theta)e^{-i(\omega t-m\varphi)},\label{eq:ansatz}
\ee
to obtain the following angular and radial equations, respectively, 
\begin{align}
\frac{1}{\sin\theta}\frac{{\d}}{{\d}\theta}\left(\sin\theta\frac{{\d}\swf_{-1\lmw}(\theta)}{{\d}\theta}\right)+\mathtt{A}_{\slmw}\swf_{-1\lmw}(\theta)=0,\label{eq:angular_eq}\\
\Delta\f{{\d}}{{\d}r}\l(\frac{{\d}\rtf_{-1\lmw}(r)}{{\d}r}\r)+\mathtt{V}_{-1\lmw}(r)\rtf_{-1\lmw}(r)=0,
\label{eq:radial_eq}
\end{align}
in which
\begin{eqnarray}
\mathtt{V}_{-1\lmw}(r) &\equiv& \f{1}{\Delta}\left[K^{2}+2i(r-M)K\right] \\ &&-\lambda_{-1\lmw}-4i\omega r, \non
\mathtt{A}_{-1\lmw}(r) &\equiv&
2a\omega\l(m+\cos\theta\r)-\f{\l(m-\cos\theta\r)^2}{\sin^{2}\theta} \non &&+\lambda_{-1\lmw}-1 - a^2 \omega^2 \sin^2 \theta,
\end{eqnarray}
where $K\equiv(r^{2}+a^{2})\omega-am$~\footnote{Eq.~(8) of Ref.~\cite{Leite:2017zyb} contains misprints. The correct expressions of $U_{\mathfrak{s}lm\omega}$ and $V_{\mathfrak{s}lm\omega}$ are $U_{\mathfrak{s}lm\omega}\equiv 2a\omega\l(m-\mathfrak{s}\cos\theta\r)-\f{\l(m+\mathfrak{s}\cos\theta\r)^2}{\sin^{2}\theta}+\lambda_{\mathfrak{s}\lmw}+\mathfrak{s} - a^2 \omega^2 \sin^2 \theta$ and $V_{\mathfrak{s}lm\omega}\equiv \f{1}{\Delta}\left[K^{2}-2i\mathfrak{s}(r-M)K\right]-\lambda_{\mathfrak{s}\lmw}+4i\mathfrak{s}\omega r$, respectively. The misprints do not affect any other expression or result in Ref.~\cite{Leite:2017zyb}.}. 
Here $\lambda_{-1\lmw}$ is the angular separation constant.

We are interested in solutions of Eq.~\eqref{eq:radial_eq} that are purely ingoing 
at the event horizon and that satisfy the following boundary conditions:
\be
\rtf_{-1\lmw}\sim\begin{cases}
\mathcal{T}_{\lmw}e^{-\imath\omtil x}\Delta , & r\rightarrow r_+,\\
\mathcal{I}_{\lmw}\f{e^{-\imath\omega x}}{r}+\mathcal{R}_{\lmw}re^{\imath\omega x}, & r\rightarrow 
+\infty,\label{eq:ingoing_sol}
\end{cases} 
\ee
where $\omtil\equiv\omega-am/2Mr_+$. The tortoise coordinate $x$ in 
Boyer-Lindquist coordinates is defined by $x\equiv\int\text{d}r\f{(r^2+a^2)}{\Delta}$, from which 
follows that $x\rightarrow+\infty$ as $r\rightarrow+\infty$; and $x\rightarrow-\infty$ as
 $r\rightarrow r_+$. 

\section{Absorption cross section} \label{sec:absorption}
We use the partial-wave method to obtain the absorption cross section. Reference~\cite{Futterman:1988ni} contains the basic steps to obtain the absorption cross section via the partial wave analysis, and here we restrict ourselves to outline the main formulas used in our computations.

The absorption cross section $\sigma_{abs}$ for an asymptotic electromagnetic plane wave travelling in the direction~$\hat{n}=\sin\gamma \,\hat{x}+\cos\gamma \, \hat{z}$ is given by
\be
\sigma_{abs}(\omega)=\frac{4\pi^{2}}{\omega^{2}}\sum_{l=1}^{+\infty}\sum_{m=-l}^{+l}\left|
\swf_{-1\lmw}(\gamma)\right|^{2}\Gamma_{\lmw},\label{eq:absorption_cs}
\ee
where $\swf_{-1\lmw}$ are the spin-weighted spheroidal harmonics for the electromagnetic case ($\mathfrak{s}=-1$), and the transmission factors $\Gamma_{\lmw}$ are given by
\be
\Gamma_{\lmw}=1-\frac{B_{lm\omega}^{2}}{16\omega^4}\left|\frac{\mathcal{R}_{lm\omega}}{ 
\mathcal{I}_{lm\omega}}\right|^{2},\label{eq:transmission_factor}
\ee
in which $B_{lm\omega}^{2}\equiv \lambda_{-1\lmw}^2+4am\omega-4a^2\omega^2$, and $\mathcal{I}_{\lmw}$, $\mathcal{R}_{\lmw}$ are the coefficients appearing in the ingoing solutions~[see Eq.~\eqref{eq:ingoing_sol}].

Here $\omega$ can be either positive or negative. We divide the incident waves according to the sign of the frequency $\omega$, into two 
categories: circularly-polarized waves co-rotating~($\omega > 0$) with the BH, and circularly-polarized waves counter-rotating~($\omega < 0$) with the BH. As we will see, co- and 
counter-rotating incident waves are absorbed in a different way. 

\section{Numerical method} \label{sec:numerical_method}
As can be seen in Eq.~\eqref{eq:absorption_cs}, the spin-weighted spheroidal harmonics $\swf_{-1\lmw}$, and the transmission factor $\Gamma_{\lmw}$ are the main ingredients to determine the absorption cross section. In this section, we detail the techniques used to obtain both $\swf_{-1\lmw}$~(Subsec.~\ref{sec:spheroidal_hamonics}) and $\Gamma_{\lmw}$~(Subsec.~\ref{sec:transmission_factor}).

\subsection{Spin-weighted spheroidal harmonics} 
\label{sec:spheroidal_hamonics}
We aim to solve the angular equation~\eqref{eq:angular_eq} in order to obtain the spin-weighted spheroidal harmonics $\swf_{-1\lmw}$, as well as the angular separation constant $\lambda_{-1lm\omega}$. The angular separation constant is needed in order to solve the radial equation~\eqref{eq:radial_eq}. There are a variety of methods available to solve the angular equation (see Ref.~\cite{berti2006eigenvalues} for a discussion). In this paper, we employ the Cook and Zalutskiy spectral decomposition approach~\cite{cook,hughes}, which is detailed in Appendix \ref{appendix:spectral}.

\subsection{Transmission factors and the Detweiler transformation} 
\label{sec:transmission_factor}
The transmission factors $\Gamma_{\lmw}$ are expressed in Eq.~(\ref{eq:transmission_factor}) in terms of the complex constants $\mathcal{I}_{\lmw}$ and $\mathcal{R}_{\lmw}$ of the radial solutions in the asymptotic regime [Eq.~(\ref{eq:ingoing_sol})] and the angular eigenvalue $\lambda_{-1lm\omega}$. 
In principle, the constants $\mathcal{I}_{\lmw}$, $\mathcal{R}_{\lmw}$ can be obtained via numerical integration of the radial equation (\ref{eq:radial_eq}). However, the different power-law fall-offs in Eq.~\eqref{eq:ingoing_sol} for the incoming ($r^{-1}$) and outgoing ($r^1$) parts poses a practical challenge for a numerical matching procedure. 
One may avoid the intrinsic complications of the `peeling' behaviour by working instead with a short-range equation that can be obtained via one of two different approaches: one due to Detweiler~\cite{detweiler1976equations} and another due to Sasaki and Nakamura~\cite{sasaki1982gravitational,sasaki2003analytic,hughes2000computing,hughes2000computing_erratum}. 

We adopt Detweiler's approach. For a full derivation of Detweiler's transformation in the Kerr-Newman background, see Ref.~\cite{i2004electromagnetic}. The radial function $\rtf_{-1\lmw}$ is transformed to a new radial function $X_{\lmw}$, which satisfies a short-range equation given by~\cite{detweiler1976equations,i2004electromagnetic}
\be
\f{{\d}^2 X_{\lmw}}{{\d} x^2}-U_{\lmw}(r)X_{\lmw} =0,\label{eq:short_range_eq}
\ee
with
\bea
&U_{\lmw}(r)\equiv-\l(\omega-\f{a
   m}{a^2+r^2}\r)^2+\f{\lambda_{-1\lmw}  \Delta}{\left(a^2+r^2\right)^2}\non&-\f{\varpi(r-M)^2\Delta\l[2 \l(\zeta^2+r^2\r)^2-\varpi\Delta\r]}{\l(a^2+r^2\r)^2\l[\l(\zeta ^2+r^2\r)^2+\varpi\Delta\r]^2}\non
   &-\f{\Delta\l[\Delta\l(2\zeta ^2+10 r^2\r)-\l(\zeta
   ^2+r^2\r)\l(\zeta ^2-10 M r+11
   r^2\right)\r]}{\left(a^2+r^2\right)^2
   \l[\l(\zeta ^2+r^2\r)^2+\varpi\Delta\r]}\non
   &+\f{12 r \Delta\l(\zeta
   ^2+r^2\r)^2 \l[r \Delta-(r-M)
   \l(\zeta
   ^2+r^2\r)\r]}{\l(a^2+r^2\r)^2
   \l[\l(\zeta ^2+r^2\r)^2+\varpi  \Delta\r]^2}
   \non
  &-\f{\Delta\l(4 a^2 M
   r+r^2 \Delta
  \r)}{\l(a^2+r^2\r)^4},
\eea
where $\vartheta\equiv2B_{\lmw}$ and
\be
\varpi\equiv\f{\vartheta -2 \lambda_{-1\lmw} }{4 \omega ^2}, \quad 
\zeta^2\equiv a\l(\f{a\omega-m}{\omega}\r).
\ee
Similarly to the long-range equation~\eqref{eq:radial_eq}, the short-range equation~\eqref{eq:short_range_eq} has a set of solutions that are purely ingoing in the event horizon and satisfy the following boundary conditions
\be
X_{\lmw}\sim\begin{cases}
\mathcal{T}^{D}_{\lmw}e^{-i \omtil x}, & r\rightarrow r_+,\\
e^{-i \omega x}+\mathcal{R}^{D}_{\lmw}e^{i \omega x}, & r\rightarrow 
+\infty,\label{eq:ingoing_sol_2}
\end{cases} 
\ee
where the superscript $D$ refers to Detweiler quantities. As pointed out by Detweiler, the Wronskian of two linearly independent solutions of Eq.~\eqref{eq:short_range_eq} is constant, so that one can show that the Detweiler coefficients satisfy the following relation
\be
\l|\mathcal{R}^{D}_{\lmw}\r|^2=1-\f{\omtil}{\omega}\l|\mathcal{T}^{D}_{\lmw}\r|^2,\label{eq:conservation_flux}
\ee
which represents the conservation of the energy flux of the incident electromagnetic plane wave. From Eq.~\eqref{eq:conservation_flux}, we note that $\l|\mathcal{R}^{D}_{\lmw}\r|^2>1$, when $\omtil\omega<0$, i.e., the wave is scattered off by the Kerr BH with an amplitude larger than the incident one. This phenomenon, known as superradiance, can lead to negative absorption cross sections in the low-frequency regime.

The Detweiler coefficient $\mathcal{R}^{D}_{\lmw}$ is related to the coefficients $\mathcal{R}_{\lmw}$ and $\mathcal{I}_{\lmw}$ appearing in Eq.~\eqref{eq:ingoing_sol} via \cite{casals2005canonical}
\be
\l|\mathcal{R}^{D}_{\lmw}\r|^2=\f{B_{\lmw}^{2}}{16\omega^4}\l|\f{\mathcal{R}_{\lmw}}{ 
\mathcal{I}_{\lmw}}\r|^{2}.\label{eq:coeffs_rel}
\ee
Thus the transmission factor of Eq.~(\ref{eq:transmission_factor}) is simply 
\be
\Gamma_{\lmw} = 1 - \l|\mathcal{R}^{D}_{\lmw}\r|^2 . \label{eq:transalt}
\ee 
An expression relating $\mathcal{T}_{\lmw}$, $\mathcal{I}_{\lmw}$, and $\mathcal{T}^{D}_{\lmw}$ can be found~\cite{casals2005canonical}, but for our current purposes the relation in Eq.~\eqref{eq:coeffs_rel} is sufficient.

To compute the transmission factors \eqref{eq:transalt}, we start by integrating Detweiler equation~\eqref{eq:short_range_eq} from a point near to the event horizon $r\sim r_+$, using a series expansion of the ingoing boundary condition given in Eq.~\eqref{eq:ingoing_sol_2}, up to a point in a region far from the BH ($r\gg r_+$), where we compare our numerical solutions with the analytical ones \eqref{eq:ingoing_sol_2}, in order to obtain Detweiler coefficients.  

\section{Numerical results}\label{sec:num_results}
In this section we present numerical results for the absorption cross section obtained using the scheme presented in Sec.~\ref{sec:numerical_method}. We  consider electromagnetic plane waves with a circular polarization impinging upon a rotating BH at an incident angle $\gamma$ (see Fig.~\ref{fig:sketch}). 

\subsection{Transmission factors}
\begin{figure*}
	\includegraphics[width=\columnwidth]{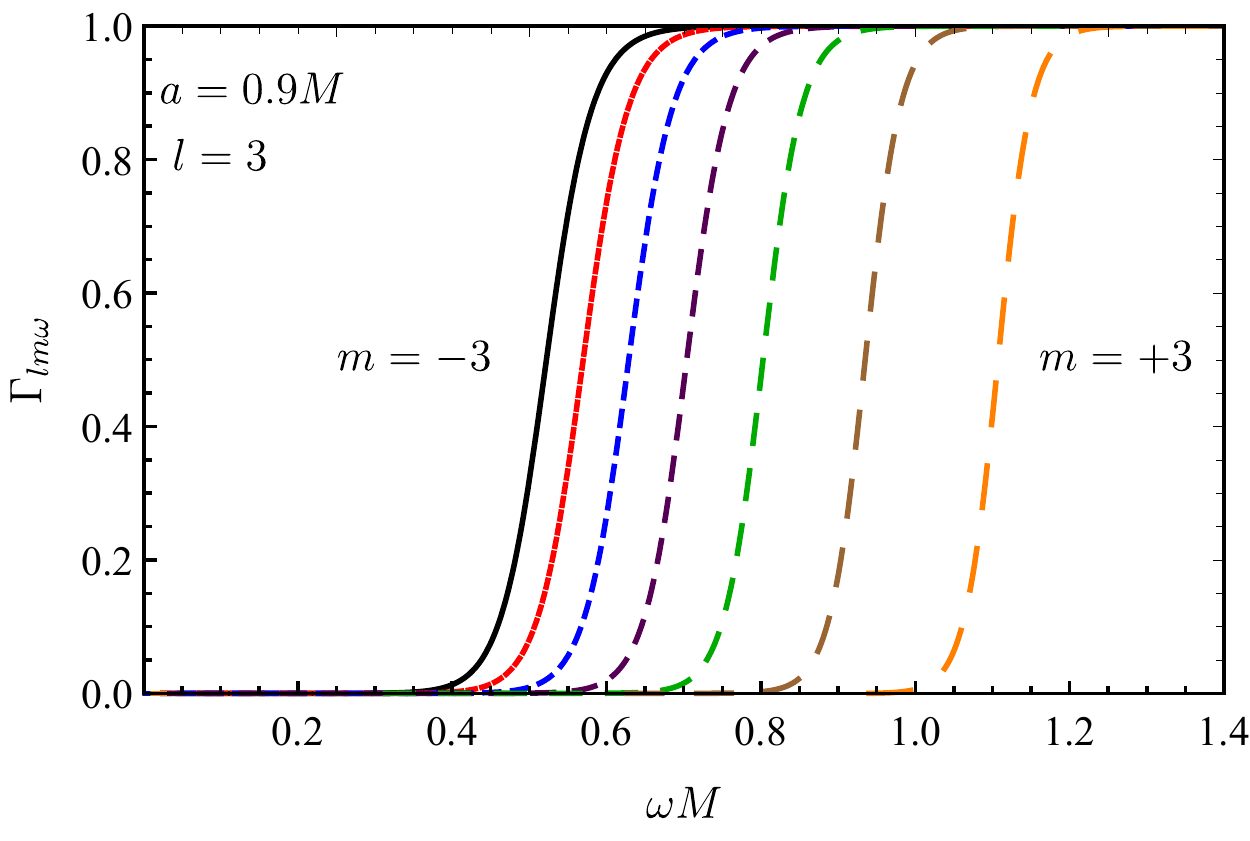}
	\caption{Transmission factors for different angular modes. The plot shows the transmission factors for the angular modes $l=3$ and $-l\leq m\leq +l$ as functions of $\omega M$, considering a Kerr BH with $a=0.9M$.
	}
	\label{fig:transfac}
\end{figure*}

Figure \ref{fig:transfac} shows typical transmission factors, calculated via Eq.~(\ref{eq:transalt}), for the angular modes $l=3$ as functions of the frequency $M \omega$. The modes are split by azimuthal number $m$. The counter-rotating modes ($m M \omega < 0$) exhibit a greater degree of transmission than the co-rotating modes ($m M \omega > 0$) at a given frequency. Results for modes with $M\omega < 0$ can be inferred from the symmetry $\Gamma_{\lmw}=\Gamma_{l(-m)(-\omega)}$. 

\subsection{Absorption cross sections: on-axis incidence}\label{subsec:on_axis_incidence}
We begin by considering the special case of waves impinging along the black 
hole rotation axis ($\gamma = 0$). The identity $S_{-1lm\omega}(0) = \delta_{m1} S_{-1l1\omega}(0)$ implies that
only partial waves with $m=1$ will contribute to the absorption cross section. This results in $\sigma_{abs}(M\omega)$ displaying a regular oscillatory pattern due to the contributions from successive partial waves in $l$. 

\begin{figure*}
\includegraphics[width=\columnwidth]{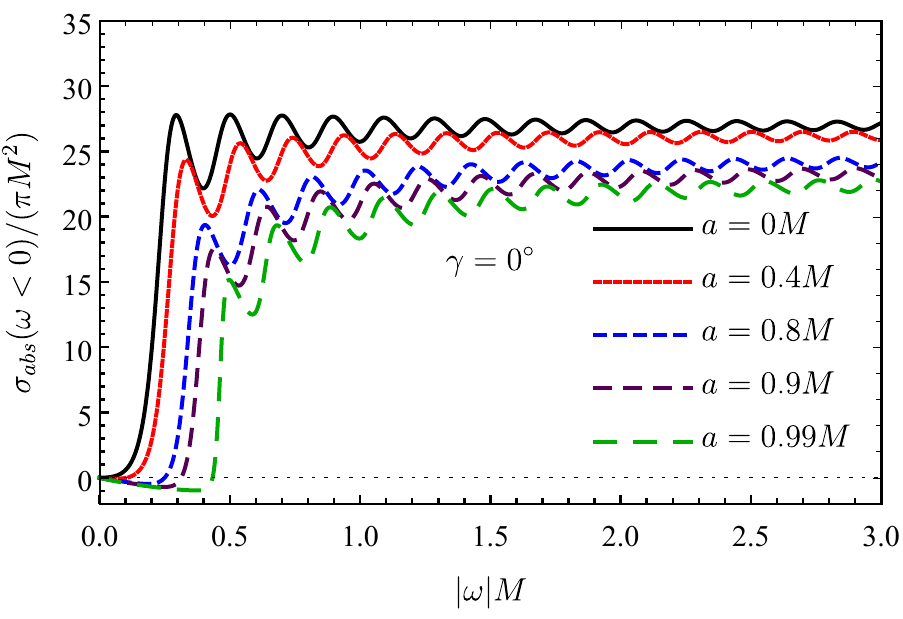}
\includegraphics[width=\columnwidth]{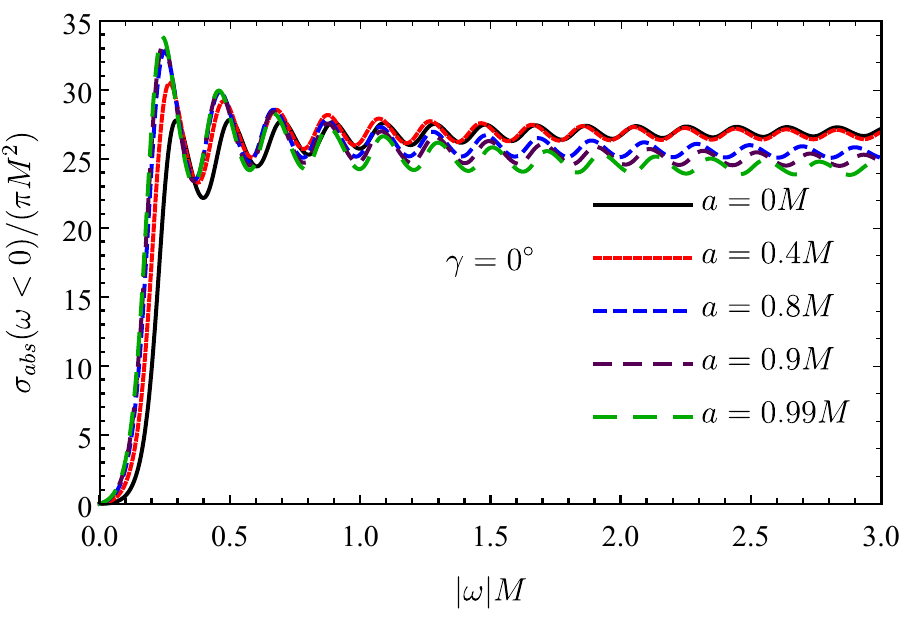}
\caption{
Absorption cross sections $\sigma_{abs}$ as a function of $M\omega$ for circularly-polarized electromagnetic waves incident parallel to the black hole rotation axis, showing the co-rotating (left) and counter-rotating (right) helicities. The five series correspond to rotation parameters $a=$~0~(Schwarzschild BH), $0.4M$, $0.8M$, $0.9M$, and $0.99M$. }
\label{fig:abs_on_axis_co_counter}
\end{figure*}

Figure \ref{fig:abs_on_axis_co_counter} shows the absorption cross sections for circularly-polarized waves which are co-rotating~(left panel, $\omega > 0$) and counter-rotating (right panel, $\omega < 0$) with the BH. 
Due to the coupling between the wave polarization and the BH rotation, co-rotating and counter-rotating incident waves are not absorbed in the same way.

For the co-rotating polarization (left panel of FIG.~\ref{fig:abs_on_axis_co_counter}), we see that superradiance [$\sigma_{abs}(\omega)<0$] occurs at low frequencies, $\omega < \Omega_H$. On the other hand, when considering counter-rotating polarizations (right panel of FIG.~\ref{fig:abs_on_axis_co_counter}) superradiance is absent, and $\sigma_{abs}$ is strictly positive across the whole frequency range. The absorption cross section tends to zero in all cases as $M\omega \rightarrow 0$, as anticipated from Eq.~(\ref{eq:lowfreq}).

In the co-rotating case, superradiance delays the onset of absorption until $\omega \approx \Omega_H$, with $\Omega_H \rightarrow 1/ 2M$ as $a \rightarrow M$. The left panel of Fig.~\ref{fig:abs_on_axis_co_counter} shows that increasing $a/M$ diminishes the absorption cross section for the co-rotating polarization, essentially monotonically. By contrast, the counter-rotating polarization [right panel of Fig.~\ref{fig:abs_on_axis_co_counter}, $\sigma_{abs}(\omega<0)$] displays a switch in behaviour as $M\omega$ increases: for $M\omega \lesssim 0.3$, the cross section is largest for rapidly-spinning BHs, whereas at high frequencies, the cross section is largest for non-spinning (Schwarzschild) BHs.  

In the short-wavelength limit $M\omega \rightarrow 0$, the absorption cross section $\sigma_{abs}$ tends towards the geodesic capture cross section $\sigma_{geo}$ (see Fig.~\ref{fig:sketch}). However, the approach to this limit depends on the helicity of the incident wave. In Ref.~\cite{Leite:2017zyb} it was shown that there is a secular term in $(\sigma_{abs} - \sigma_{geo})$ that scales with $- a s (M \omega)^{-1}$ in the short-wavelength regime. This term has the opposite sign to $\omega$, as more of the counter-rotating polarization is absorbed than the co-rotating polarization. This is an example of a spin-helicity effect \cite{Leite:2017zyb}, associated with a coupling between wave helicity and the frame-dragging of spacetime.

\begin{figure*}
\includegraphics[width=1.4\columnwidth]{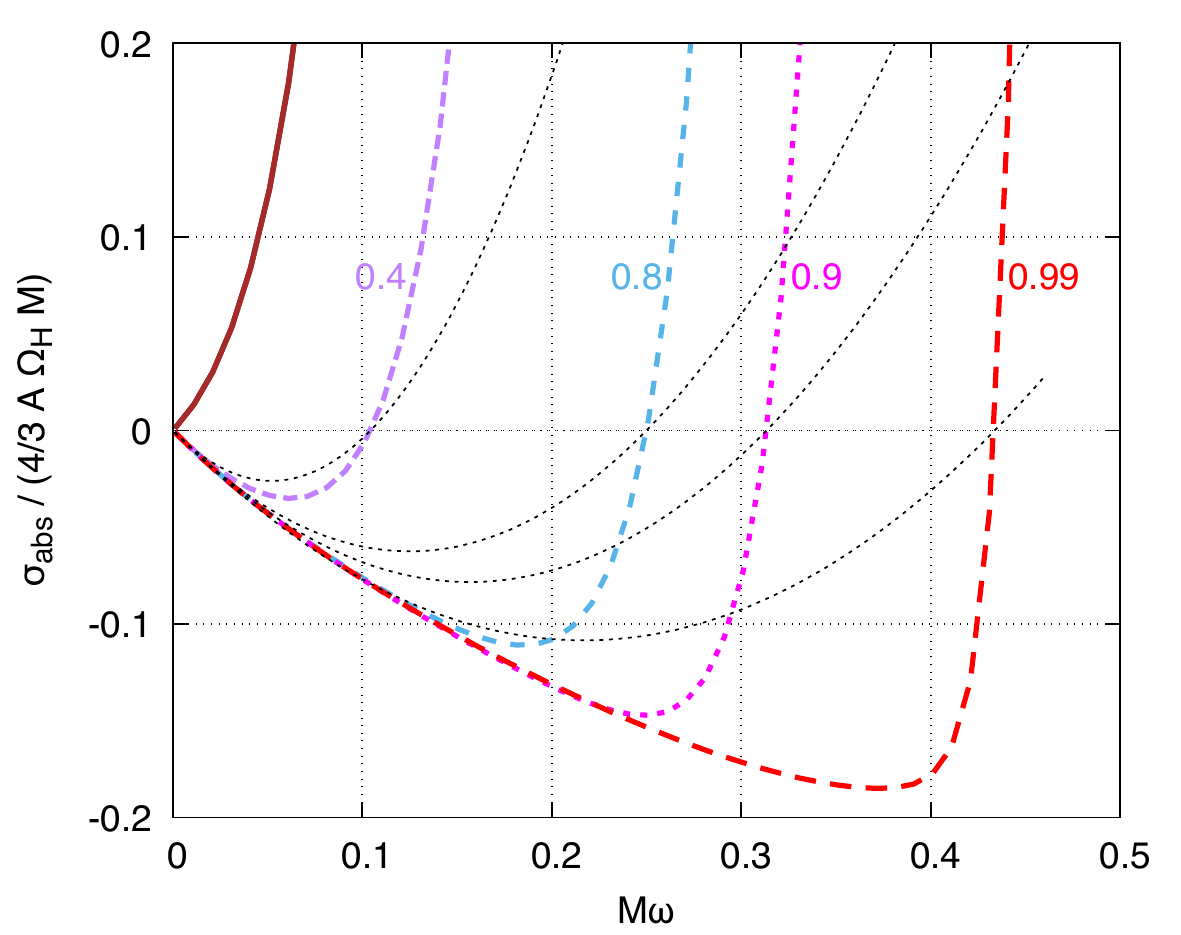}
\caption{Electromagnetic absorption at low frequencies. The plot shows the absorption cross section $\sigma_{abs}$ divided by $\frac{4}{3} \mathcal{A} \Omega_H M$, where $\mathcal{A}$ and $\Omega_H$ are the area and angular frequency of the event horizon, respectively [cf.~Eq.~(\ref{eq:lowfreq})]. The thick dashed lines are for the co-rotating polarization, for BH spin ratios of $a = 0.4M$, $a = 0.8M$, $a = 0.9M$ and $a=0.99M$. The thick solid lines, which overlie each other, are for the counter-rotating polarization for the same values of $a$. The thin dotted lines show the approximations $- \omega (1 - \omega / \Omega_H)$. 
 }
\label{fig:lowfreq}
\end{figure*}

Figure \ref{fig:lowfreq} shows the electromagnetic absorption cross sections at low frequencies, in the on-axis case. The plot confirms that the long-wavelength approximation of Eq.~(\ref{eq:lowfreq}) successfully describes the leading-order scaling at low frequencies, $\sigma_{abs} \sim - \frac{4}{3} \mathcal{A} \Omega_H \omega M^2$. Eq.~(\ref{eq:lowfreq}) also correctly anticipates the change of sign in $\sigma_{abs}$ at the superradiant frequency $\omega = \Omega_H$. However, in the intermediate regime $0 < \omega < \Omega_H$, Fig.~\ref{fig:lowfreq} shows that Eq.~(\ref{eq:lowfreq}) somewhat underestimates the magnitude of superradiant emission.

\subsection{Absorption cross sections: off-axis incidence}\label{subsec:off_axis_incidence}
Now we turn to the general case in which the incident wave's propagation direction is not aligned with the BH spin axis. In this section we will consider the following choices of the rotation parameter: $a/M=0.4$, $0.8$, $0.9$, and $0.99$. The cross section becomes more irregular as the angle of incidence increases, due to competition between transmission factors of the same $l$ but differing $m$. Whereas in the Schwarzschild case $m$-modes with the same $l$ are degenerate, in the Kerr case the $m$-modes are split by rotation (see Fig.~\ref{fig:transfac}). For $\gamma \neq 0^\circ$, all the $m$ modes (not just $m=1$, as in the $\gamma = 0$ case) contribute to the mode sum in Eq.~(\ref{eq:absorption_cs}).

\begin{figure*}
\includegraphics[width=0.94\columnwidth]{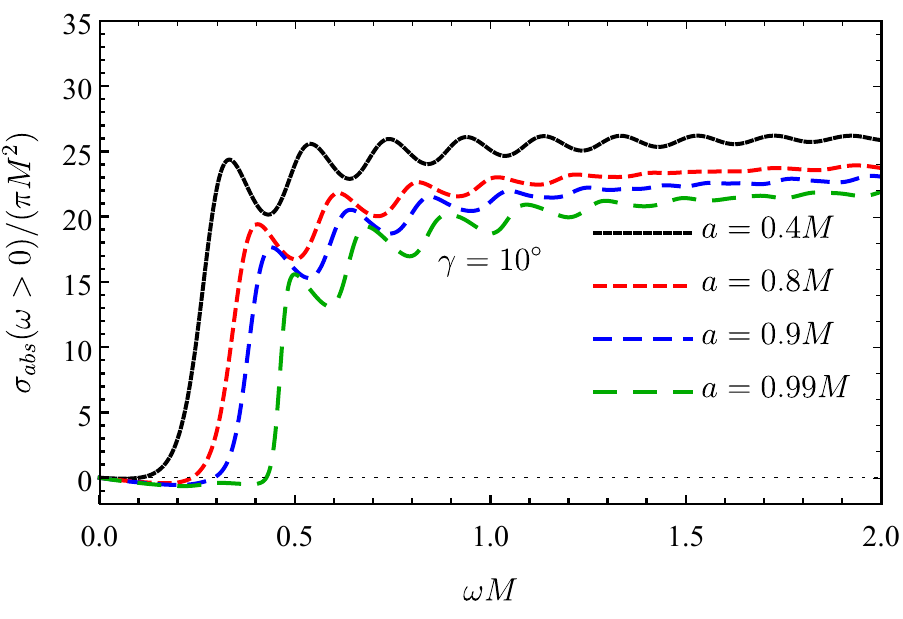}  
\includegraphics[width=0.94\columnwidth]{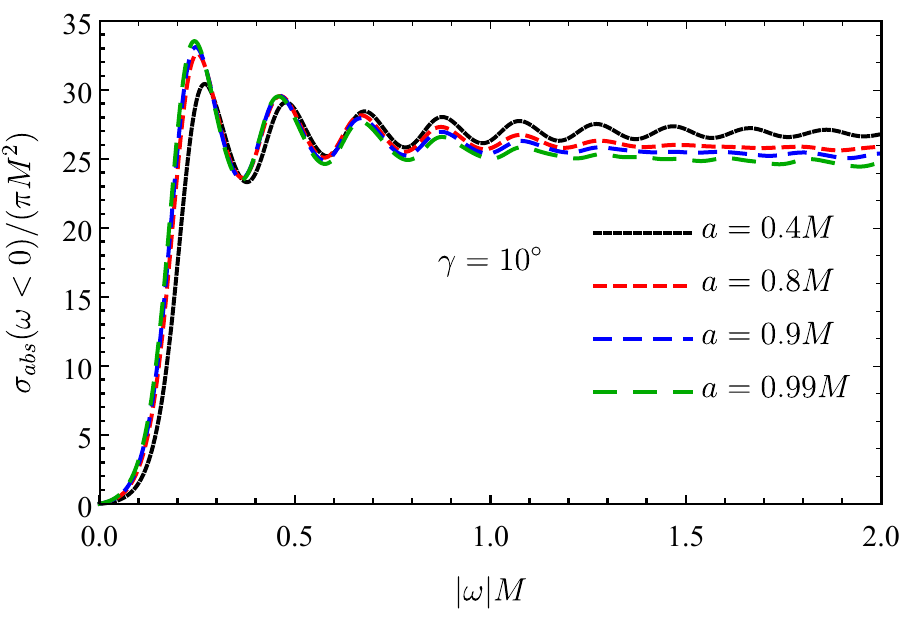} \\
\includegraphics[width=0.94\columnwidth]{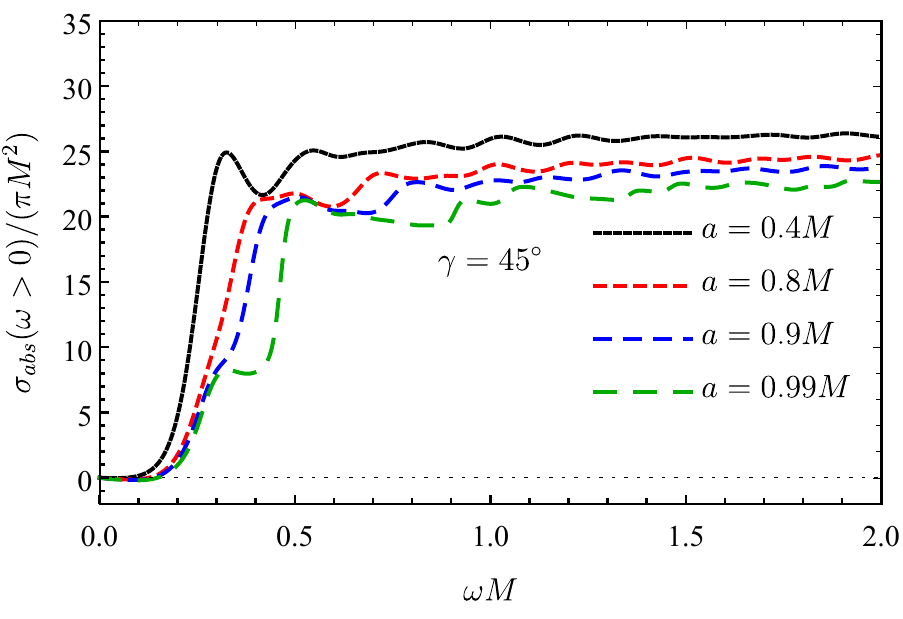}
\includegraphics[width=0.94\columnwidth]{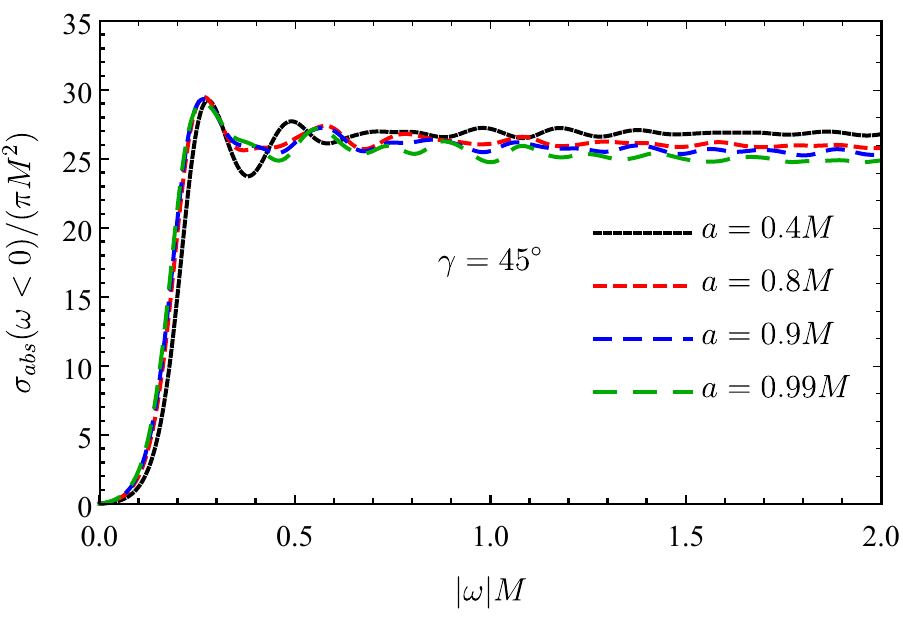} \\
\includegraphics[width=0.94\columnwidth]{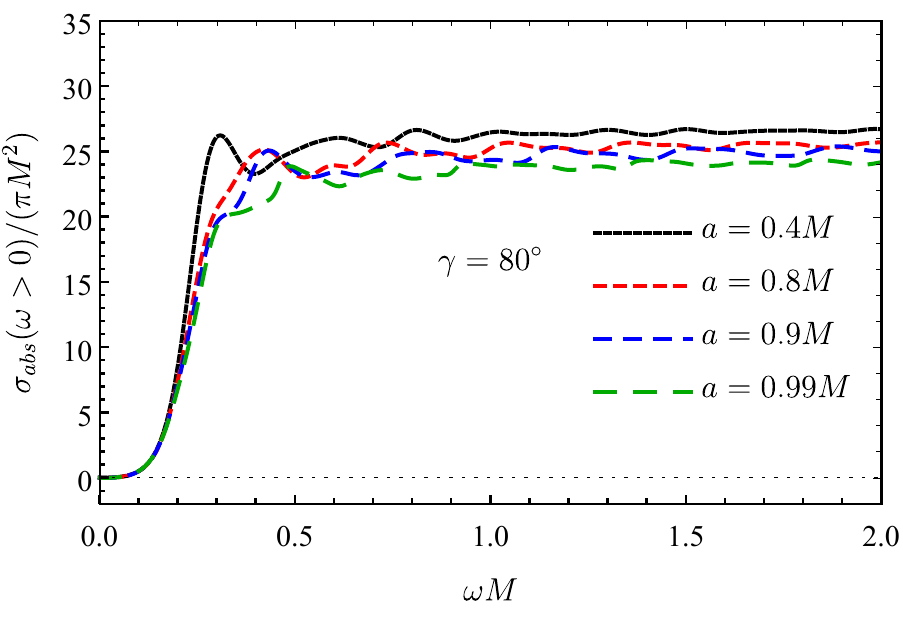} 
\includegraphics[width=0.94\columnwidth]{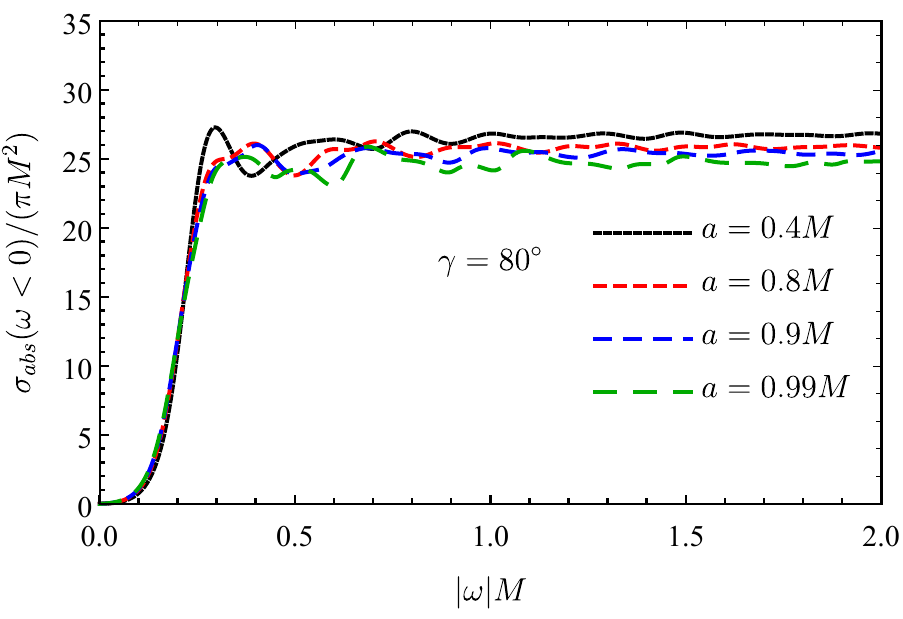} \\
\includegraphics[width=0.94\columnwidth]{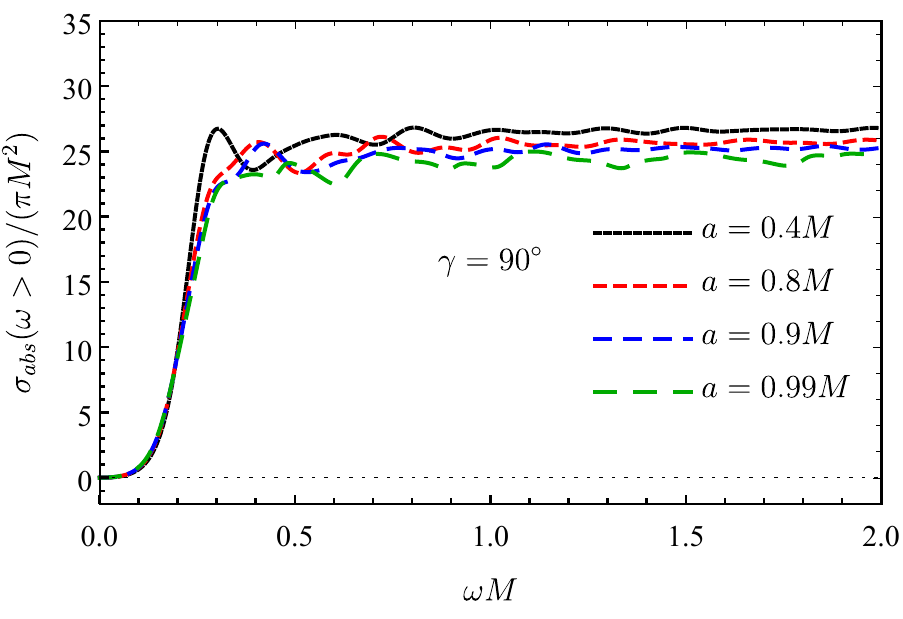}
\includegraphics[width=0.94\columnwidth]{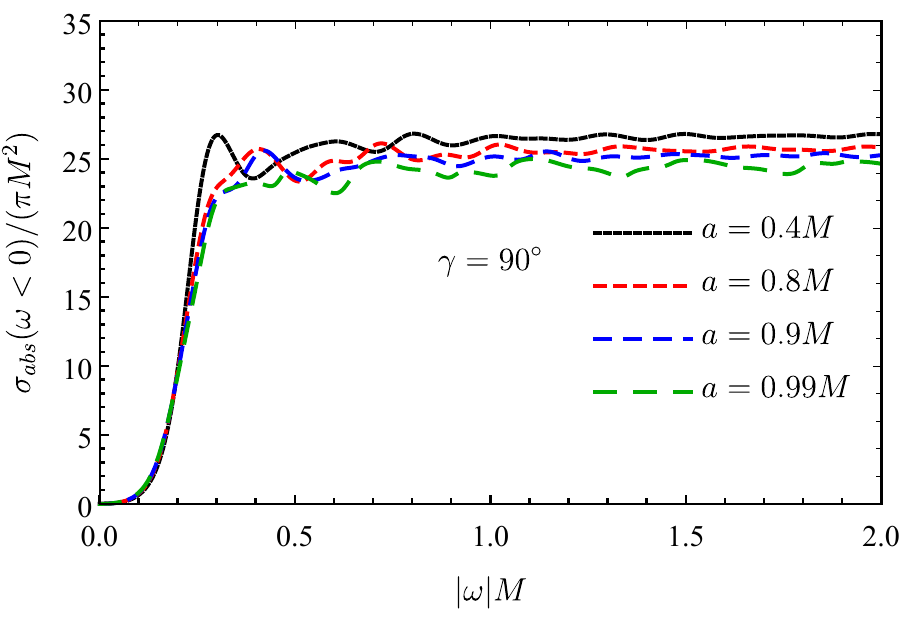}
\caption{Absorption cross sections for co-rotating (left) and counter-rotating (right) circular polarizations, for angles of incidence $\gamma = 10^\circ$, $45^\circ$, $80^\circ$ and $90^\circ$. 
The cross section becomes more irregular as the angle of incidence increases.   In the case of equatorial incidence ($\gamma = 90^\circ$) there is no difference between the cross sections of the two circular polarizations (co- and counter-rotating).
}
\label{fig:abs_off_axis}
\end{figure*}

Figure~\ref{fig:abs_off_axis} shows the off-axis cross sections for the co-rotating polarization (left) and counter-rotating polarization (right). In general, $\sigma_{abs} /M^2$ is smaller for the former than for the latter; and it decreases in both cases as $a/M$ increases. For small $\gamma$ (i.e.~$\gamma = 10^\circ$), the cross section profile resembles the on-axis case. However, as the incidence angle increases from $\gamma=0^{\circ}$ to $\gamma=90^\circ$, the absorption cross section becomes less regular. Contributions from a multitude of partial waves overlap, and thus the oscillations are less distinct. This irregular behavior presented by the absorption cross sections for off-axis incidences, is a consequence of 
the coupling between the BH rotation and the wave azimuthal number~$m$~\cite{glampedakis2001scattering,Caio:2013}. A similar behaviour was found in the scalar-field case (see Fig.~$3$ in Ref.~\cite{Caio:2013}).

\subsection{Superradiance}\label{subsec:superradiance}
We now examine superradiance in the off-axis case.
Superradiance ($\Gamma < 0$) occurs in all multipole modes satisfying the condition $\omtil \omega < 0$, which is only satisfied for co-rotating incident waves. For on-axis ($\gamma = 0$) co-rotating ($\omega > 0$) planar waves, net superradiance occurs whenever $a\neq0$~(left panel in FIG.~\ref{fig:abs_on_axis_co_counter}). For off-axis waves the situation is more delicate.

\begin{figure*}
\includegraphics[width=\columnwidth]{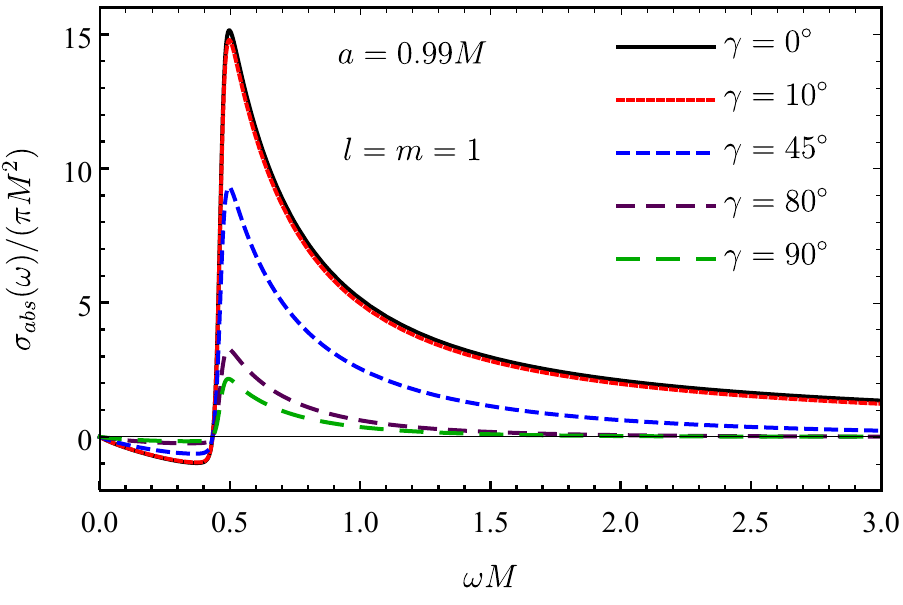}
\caption{Partial absorption cross sections~($l=m=1$) for different incidence angles~($\gamma=0^{\circ}$, $10^{\circ}$, $45^{\circ}$, $80^{\circ}$, and $90^{\circ}$) and $a=0.99M$. Superradiance is larger for on-axis incidence.}
\label{fig:partial_omeplus_lm1}
\end{figure*}

By far the largest superradiant contribution comes from the co-rotating dipole mode $l=1$, $m=1$ mode, and superradiance is exponentially suppressed in the higher multipoles. Figure \ref{fig:partial_omeplus_lm1} shows the partial absorption cross section for the $l=m=1$ mode, for the cases $\gamma=0^{\circ}$, $10^{\circ}$, $45^{\circ}$, $80^{\circ}$, and $90^{\circ}$, for  $a=0.99M$. As previously emphasized, superradiance is larger for a wave impinging for a direction parallel to the rotation axis, than for a wave from a direction in the the equatorial plane. 

For off-axis incidences, superradiance weakens as one deviates the incidence angle away from the BH rotation axis. At the equatorial plane~($\gamma=90^{\circ}$, bottom right panel of FIG.~\ref{fig:abs_off_axis}) 
the total absorption cross section is non-negative even at low frequencies $\omega$. 
This is a consequence of the fact that the off-axis cross section is a sum of all the multipoles~($l$, $m$), so that we need to take into account the contributions 
of both superradiant~($m\geq1$) and non-superradiant~($m<1$) modes. As $\gamma$ approaches $\pi/2$, the 
competition between superradiant and non-superradiant modes
is won by the latter, so that, close/along the equatorial plane, the emission in the superradiant mode is less than the  absorption is the non-superradiant modes, leading to an overall net absorption~\cite{rosa2015testing}.

\subsection{Linearly-polarized waves}\label{subsec:linear_polarization}
We now turn our attention to the absorption of linearly-polarized incident waves, which are constructed from a linear combination of two circularly-polarized waves. The cross section in the linearly-polarized case is simply the average of the cross section for $+|\omega|$ and $-|\omega|$.

\begin{figure*}
\includegraphics[width=\columnwidth]{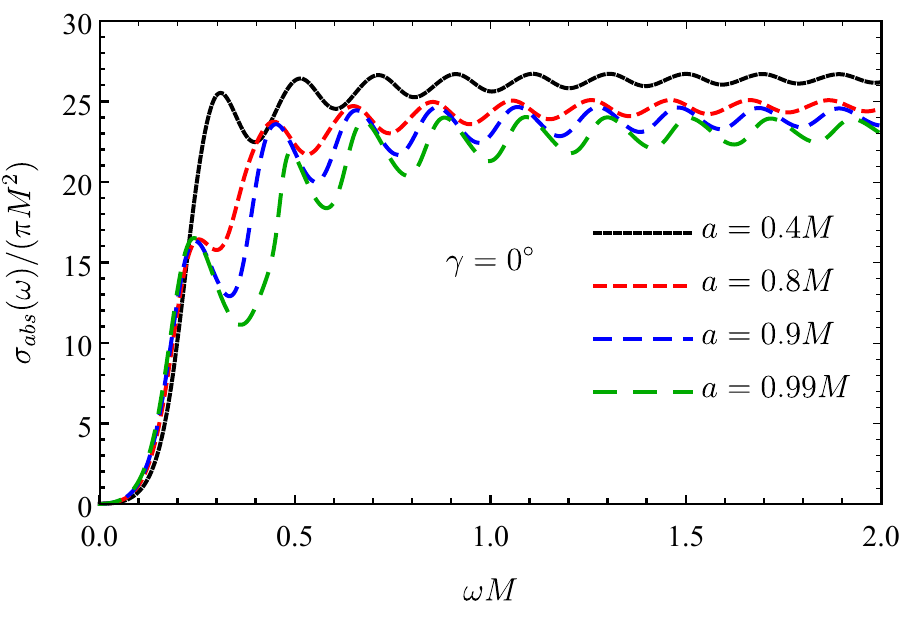}
\includegraphics[width=\columnwidth]{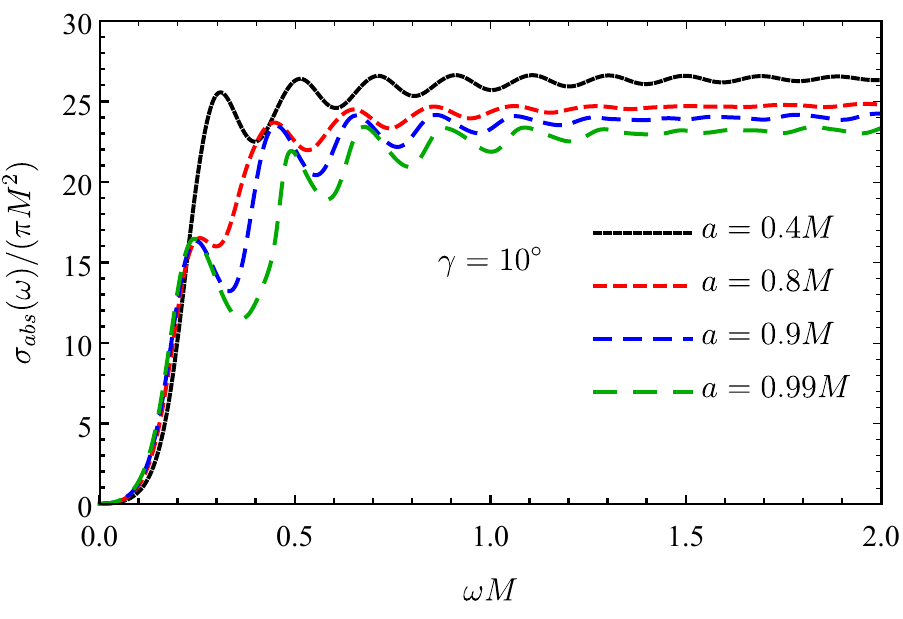}
\includegraphics[width=\columnwidth]{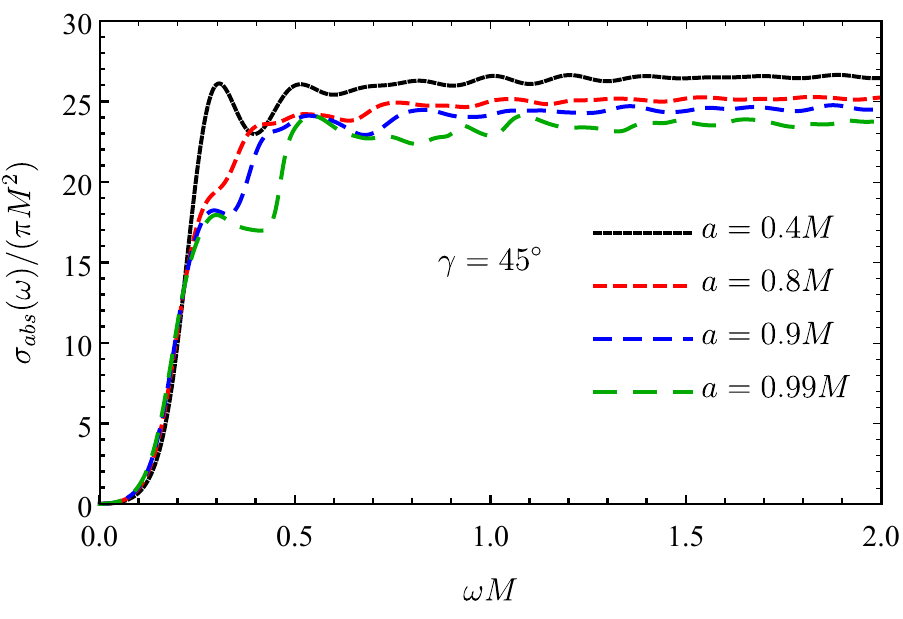}
\includegraphics[width=\columnwidth]{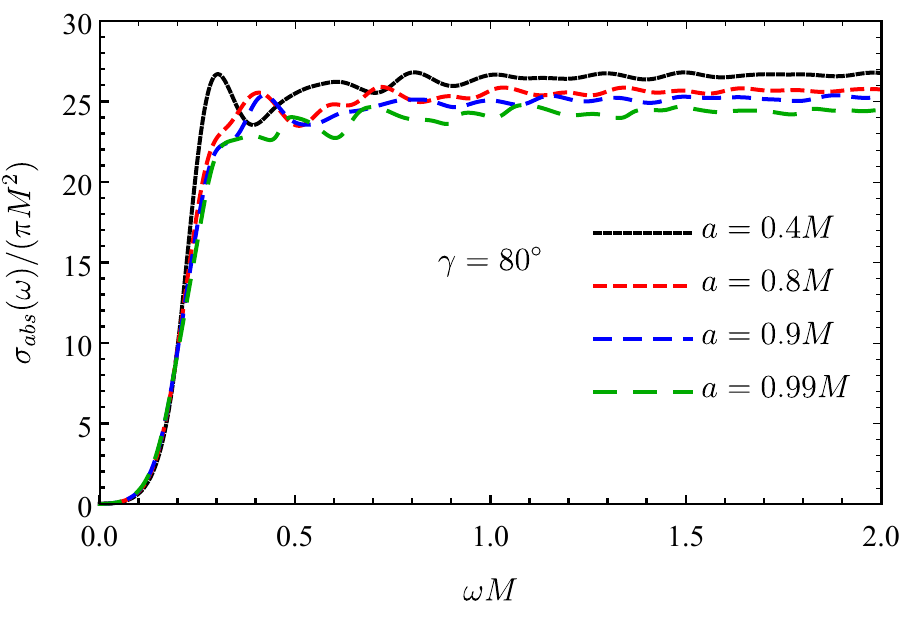}
\caption{Absorption cross sections for linearly-polarized electromagnetic waves. Absorption is larger 
the smaller is the value of the BH rotation parameter. As well as in the case of circularly-polarized waves, the absorption cross sections present a less regular behavior for off-axis incidences.}
\label{fig:linear_polarization}
\end{figure*}

Figure \ref{fig:linear_polarization} shows the absorption cross sections for linearly-polarized incident waves, 
considering different values for the incidence angle~($\gamma=0^{\circ}$, $10^{\circ}$, $45^{\circ}$, and $80^{\circ}$) and 
for the BH rotation parameter~($a/M=0.4$, $0.8$, $0.9$, and $0.99$). 
In general, the dimensionless quantity $\sigma_{abs} / M^2$ decreases as the BH spin $a/M$ increases. 
Moreover, the absorption cross sections exhibit a 
regular oscillatory behavior (with $M\omega$) for on-axis incidence, but a far less regular structure as the incidence angle increases. For the case of equatorial incidence~($\gamma=90^{\circ}$), the linearly-polarized cross section is exactly equal to the circularly-polarized cross sections, as the spin of the black hole does not distinguish between right- and left-handed polarizations. 

\begin{figure*}
\includegraphics[width=\columnwidth]{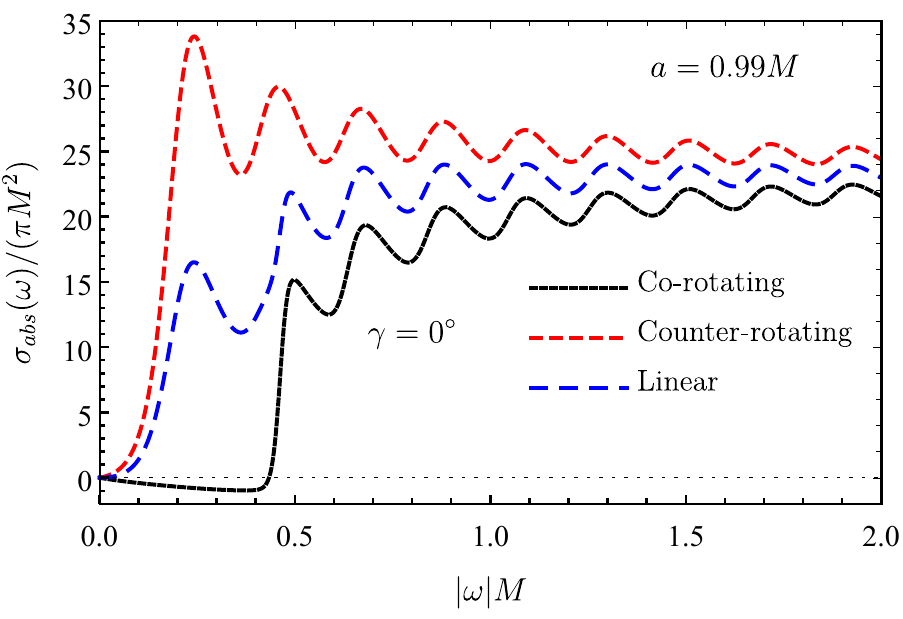}
\includegraphics[width=\columnwidth]{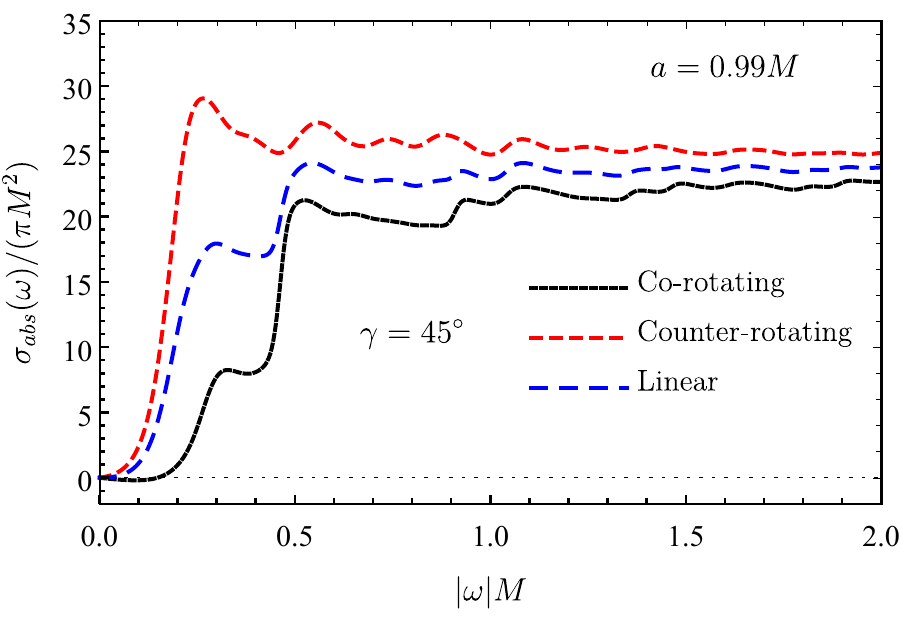}
\caption{Comparison of the results for linearly, co-rotating, and counter-rotating polarized incident electromagnetic waves.}
\label{fig:linear_co_counter_linear}
\end{figure*}

Figure \ref{fig:linear_co_counter_linear} compares the results for linearly- and circularly-
polarized incident waves. We choose two values for the incidence angle, namely, $\gamma=0^{\circ}$, and 
$45^{\circ}$; and we consider a rapidly-rotating BH, with $a=0.99M$. 
Linearly-polarized waves are less absorbed than circularly the counter-rotating waves and more absorbed than co-rotating ones. 
The cross section $\sigma_{abs}$ remains positive in the linear case, and there is no net superradiance.

\section{Final remarks}\label{sec:remarks}
We have computed the electromagnetic absorption cross section of 
Kerr BHs using numerical techniques. We have considered polarized electromagnetic plane 
waves impinging at different incidence angles upon Kerr BHs, showing that the 
absorption spectrum presents a rich structure which is directly related to the wave~(polarization, 
incidence angle, frequency, and angular momentum) and BH~(rotation) characteristics.  

We have computed the spin-weighted spheroidal harmonics and 
the angular separation constant using the spectral method proposed in Ref.~\cite{cook}.
The transmission factors have been obtained with the aid of Detweiler's formalism, as detailed 
in Sec.~\ref{sec:spheroidal_hamonics}. 
We have tested the eigenvalues~$\Lambda_{-1\lmw}$ 
against the low-$a\omega$ formula given in Ref.~\cite{berti2006eigenvalues}, 
obtaining great agreement. 
Moreover, we have checked our numerical values for the Detweiler 
coefficients~$\mathcal{R}^{D}_{\lmw}$ against the ones shown in Ref.~\cite{i2004electromagnetic}, 
obtaining concordance.

For on-axis incident waves, we have seen that the cross sections present a 
regular oscillatory behavior similar to the ones presented for static 
BHs~\cite{Crispino:2007qw,Crispino:2008zz}. 
For off-axis incidences, we have shown that the absorption cross sections present a less regular pattern than 
that presented in the on-axis case~(see Sec.~\ref{subsec:off_axis_incidence}). The irregular 
behavior is a consequence of the coupling between 
the BH angular momentum and the wave angular momentum. Many of the features presented by the 
electromagnetic absorption cross section are shared 
with the gravitational absorption cross section~\cite{Dolan:2008kf}.

For Kerr BHs, counter-rotating polarizations are more absorbed than co-rotating polarizations, due to the coupling between the wave helicity and the BH spin. This implies that a net polarization will be generated in the scattered flux, even if the incident wave is unpolarized. 

At low frequencies $\omega < \Omega_H$, the electromagnetic absorption cross section for circularly-polarized planar waves can become negative, due to superradiance. For this to occur, the wave helicity and the BH must share the same handedness. This implies that, in principle, an incoming wave can lead to stimulated emission by the BH. One interesting possibility, investigated in Ref.~\cite{Rosa:2016bli}, is that net superradiance could be excited by a pulsar orbiting a rotating black hole.

\begin{acknowledgments}
The authors would like to thank Conselho Nacional de Desenvolvimento Cient\'ifico e Tecnol\'ogico (CNPq) and Coordena\c{c}\~ao de Aperfei\c{c}oamento de Pessoal de N\'ivel Superior (CAPES), from Brazil, for partial financial support.
This research has also received funding from the European Union's Horizon 2020 research and innovation programme under the H2020-MSCA-RISE-2017 Grant No. FunFiCO-777740.
L.L.~acknowledges the School of Mathematics and Statistics of the University of Sheffield for the kind hospitality while part of this work was undertaken.
S.D.~acknowledges additional financial support from the Science and Technology Facilities Council (STFC) under Grant No.~ST/L000520/1. 
\end{acknowledgments}
%

\appendix

\section{Spin-weighted spheroidal harmonics: spectral decomposition\label{appendix:spectral}}

The differential equation obeyed by the spin-weighted
spheroidal function $\swf_{\slmw}$ can be rewritten as 
\bea
\l[ 
\l(1-u^{2}\r)\l(\swf_{\slmw}(u)\r)_{,u}\r] 
_{,u}+\Lambda_{\slmw}\swf_{\slmw} \non
+\left[\left(\aw u\right)^{2}-2\mathfrak{s}\aw u+\mathfrak{s}-\frac{
\left(m+\mathfrak{s}u\right)^{2}}{1-u^{2}}\right]\swf_{\slmw}(u)=0,
\label{eq:spheroidal_eq}
\eea
where $\Lambda_{\slmw}\equiv\lambda_{\slmw}-a^2\omega^2+2am\omega$, $u\equiv\cos\theta$, 
and $\aw\equiv a\omega$.

In the spectral decomposition approach, the spin-weighted spheroidal harmonics are expanded
in terms of the spin-weighted spherical harmonics in the following way \cite{cook}:
\begin{equation}
\swf_{\slmw}(u)=\sum_{j=l_{min}}b_{jm\omega}^{(l)}Y_{{\sw}jm}(u),
\label{eq:spectral_decomp}
\end{equation}
where $l_{min}\equiv 
\text{max}\left(\left|m\right|,\,\left|\sw\right|\right)$, 
$b_{jm\omega}^{(l)}$ are the expansion coefficients, and $Y_{{\sw}jm}$ 
represent the spin-weighted spherical harmonics. The coefficients 
$b_{jm\omega}^{(l)}$ can be calculated via a recursive formula that 
is obtained as follows.

Inserting Eq.~\eqref{eq:spectral_decomp} into 
Eq.~\eqref{eq:spheroidal_eq},
\begin{eqnarray}
\sum_{j=l_{min}}b_{jm\omega}^{(l)}\left\{ \l[ 
\left(1-u^{2}\right)Y_{{{\sw}lm}_{,u}}\r] 
_{,u}+\Lambda_{\slmw}Y_{{\sw}jm}\right.\non
+\left.\left[\left(\aw u\right)^{2}-2{\sw}\aw u+{\sw}-\frac{
\left(m+{\sw}u\right)^{2}}{1-u^{2}}\right]Y_{{\sw}jm}\right\}= 0,
\end{eqnarray}
and using the fact that $\Lambda_{{\sw}lm0}=l(l+1)-{\sw}({\sw}+1)$, we arrive at
\bea
\sum_{j=l_{min}}b_{jm\omega}^{(l)}\left[-\left(\aw u\right)^{2}+2{\sw}\aw u-\Lambda_{
\slmw}\r.\non
\l.+j(j+1)-{\sw}({\sw}+1)\right]Y_{{\sw}jm}=0.\label{eq:coeff_eq}
\eea
Since the angular dependence is carried by $Y_{{\sw}jm}$~[see Eq.~\eqref{eq:spectral_decomp}], 
we need to eliminate the $x$ dependence 
in Eq.~\eqref{eq:coeff_eq}. 
To do so, we use the
following recurrence relation for the spin-weighted 
spherical harmonics \cite{blanco} 
\bea
u Y_{{\sw}jm}&=&\mathcal{Y}_{{\sw}jm}Y_{{\sw}(j+1)m}+\mathcal{Z}_{{\sw}jm}
Y_{{\sw}(j-1)m}\non&+&\mathcal{Q}_{sjm}Y_{{\sw}jm},
\label{eq:recurrence_eq}
\eea
where we can identify
\begin{eqnarray}
\mathcal{Y}_{{\sw}jm} & = & 
\sqrt{\frac{\left(j+1\right)^{2}-m^{2}}{
\left(2j+1\right)\left(2j+3\right)}}\sqrt{\frac{\left(j+1\right)^{2}
-{\sw}^{2}}{\left(j+1\right)^{2}}},\\
\mathcal{Z}_{{\sw}jm} & = & \begin{cases}
\sqrt{\frac{j^{2}-{\sw}^{2}}{j^{2}}}\sqrt{\frac{j^{2}-m^{2}}{4j^{2}-1}}, & 
\text{for}\: j\neq0,\\
0, & \text{for}\: j=0,
\end{cases}\\
\mathcal{Q}_{{\sw}jm} & = & \begin{cases}
-\frac{m{\sw}}{j\left(j+1\right)}, & 
\text{for}\: j\neq0\:\text{and}\: {\sw}\neq0,\\
0, & \text{for}\: j=0\:\text{or}\: {\sw}=0.
\end{cases}
\end{eqnarray}

From Eq.~\eqref{eq:recurrence_eq}, we can also obtain
\begin{eqnarray}
u^{2}Y_{{\sw}jm} = 
\mathcal{Y}_{{\sw}jm}\left(\mathcal{Q}_{{\sw}(j+1)m}+\mathcal{Q}_{{\sw}jm}
\right)Y_{{\sw}(j+1)m}\non
+\mathcal{Z}_{{\sw}jm}\left(\mathcal{Q}_{{\sw}(j-1)m}
+\mathcal{Q}_{{\sw}jm}\right)Y_{{\sw}(j-1)m}\non 
+\left(\mathcal{Y}_{{\sw}jm}\mathcal{Y}_{{\sw}(j+1)m}\right)Y_{{\sw}(j+2)m}\non
+\left(\mathcal{Z}_{{\sw}jm}\mathcal{Z}_{{\sw}(j-1)m}\right)Y_{{\sw}(j-2)m}\non
+\left(\mathcal{Y}_{{\sw}jm}\mathcal{Z}_{{\sw}(j+1)m}+\mathcal{Z}_{
{\sw}jm}\mathcal{Y}_{{\sw}(j-1)m}+\mathcal{Q}_{{\sw}jm}^{2}\right)Y_{{\sw}jm}. 
\label{recurrence_eq_2}
\end{eqnarray}

Using Eqs.~\eqref{eq:recurrence_eq} and \eqref{recurrence_eq_2}
together with Eq.~\eqref{eq:coeff_eq}, we are left with
\begin{align}
\sum_{j=l_{min}}b_{jm\omega}^{(l)}\l\{\left[2\aw{\sw}\mathcal{Q}_{{\sw}jm}-\Lambda_{\slmw}
+j(j+1)-{\sw}({\sw}+1)\right.\r.\non
\left.-\aw^{2}\left(\mathcal{Y}_{{\sw}jm}\mathcal{Z}_{{\sw}(j+1)m}+\mathcal{Z}_{
{\sw}jm}\mathcal{Y}_{{\sw}(j-1)m}+\mathcal{Q}_{{\sw}jm}^{2}\right)\right]Y_{{\sw}jm}\non
+\left[2{\sw}\aw\mathcal{Y}_{{\sw}jm}-\aw^{2}\mathcal
{Y}_{{\sw}jm}\left(\mathcal{Q}_{{\sw}(j+1)m}+\mathcal{Q}_{{\sw}jm}\right)\right]Y_
{{\sw}(j+1)m}\non
+\left[2{\sw}\aw\mathcal{Z}_{{\sw}jm}-\aw^{2}\mathcal{Z}_{{\sw}jm}
\left(\mathcal{Q}_{{\sw}(j-1)m}+\mathcal{Q}_{{\sw}jm} \right)\right]Y_{{\sw}(j-1)m}\non
-\aw^{2}\left(\mathcal{Y}_{{\sw}jm}
\mathcal{Y}_{{\sw}(j+1)m}\right)Y_{{\sw}(j+2)m} \non
\l.-\aw^{2}\mathcal{Z}_{{\sw}jm}\mathcal{Z}_{{\sw}(j-1)m}Y_{{\sw}(j-2)m}\r\}= 0.
\end{align}

After some algebraic manipulations and noting that $\mathcal{Y}_{{\sw}(l_{min}-1)m}=Y_{{\sw}(l_{min}-1)m}=Y_{{\sw}(l_{min}-2)m}=0$, we can obtain a five-term recurrence relation for $b_{jm\omega}^{(l)}$:
\begin{align}
b_{jm\omega}^{(l)}\left[2\aw{\sw}\mathcal{Q}_{{\sw}jm}+j(j+1)-{\sw}({\sw}+1)\right.\nonumber\\
\left.-\aw^{2}\left(\mathcal{Y}_{{\sw}jm}\mathcal{Z}_{{\sw}(j+1)m}+\mathcal{Z}_{
{\sw}jm}\mathcal{Y}_{{\sw}(j-1)m}+\mathcal{Q}_{{\sw}jm}^{2}\right)\right]\nonumber\\
+b_{(j-1)m\omega}^{(l)}\left[2{\sw}\aw\mathcal{Y}_{{\sw}(j-1)m}-\aw^{2}\mathcal{Y}_{
{\sw}(j-1)m}\left(\mathcal{Q}_{{\sw}jm}+\mathcal{Q}_{{\sw}(j-1)m}\right)\right]\nonumber \\
+b_{(j+1)m\omega}^{(l)}\left[2{\sw}\aw\mathcal{Z}_{{\sw}(j+1)m}-\aw^{2}\mathcal{Z}_{
{\sw}(j+1)m}\left(\mathcal{Q}_{{\sw}jm}+\mathcal{Q}_{{\sw}(j+1)m}\right)\right]\nonumber \\
-\aw^{2}b_{(j-2)m\omega}^{(l)}\left(\mathcal{Y}_{{\sw}(j-2)m}\mathcal{Y}_{{\sw}(j-1)m
}\right)\non
-\aw^{2}b_{(j+2)m\omega}^{(l)}\left(\mathcal{Z}_{{\sw}(j+2)m}\mathcal{Z}_
{{\sw}(j+1)m}\right)=\Lambda_{\slmw}b_{jm\omega}^{(l)}.\label{eq:five_term_recurence_relation}
\end{align}

We can rewrite Eq.~\eqref{eq:five_term_recurence_relation} as
an eigenvalue equation, namely
\begin{equation}
X\cdot\vec{b}_{jm\omega}=\Lambda_{\slmw}\vec{b}_{jm\omega},\label{eq:eigensystem}
\end{equation}
where the coefficients $b_{jm\omega}^{(l)}$ are the elements of the eigenvectors matrix $\vec{b}_{jm\omega}$, 
and $\Lambda_{\slmw}$ are the elements of the eigenvalue matrix. 
The non-zero elements of the matrix $X$ are given by

\begin{flalign}
X_{k(k-2)}=-\aw^{2}\mathcal{Y}_{{\sw}(k-2)m}\mathcal{Y}_{{\sw}(k-1)m}&\non
X_{k(k-1)}=2{\sw}\aw\mathcal{Y}_{{\sw}(k-1)m}-\aw^{2}\mathcal{Y}_{{\sw}(k-1)m}\left(\mathcal{Q}_{{\sw}km}+\mathcal{Q}_{
{\sw}(k-1)m}\right)&\non
X_{kk}=\left[2\aw{\sw}\mathcal{Q}_{{\sw}km}+k(k+1)-{\sw}({\sw}+1)\right.\non
\left.-\aw^{2}\left(\mathcal{Y}_{{\sw}km}\mathcal{Z}_{{\sw}(k+1)m}+\mathcal{Z}_{{\sw}km}\mathcal{Y}_{
{\sw}(k-1)m}+\mathcal{Q}_{{\sw}km}^{2}\right)\right]&\non
X_{k(k+1)}=2{\sw}\aw\mathcal{Z}_{{\sw}(k+1)m}&\non
-\aw^{2}\mathcal{Z}_{{\sw}(k+1)m}\left(\mathcal{Q}_{{\sw}km}+\mathcal{Q}_{{\sw}(k+1)m}\right)&\non
X_{k(k+2)}=-\aw^{2}\mathcal{Z}_{{\sw}(k+2)m}\mathcal{Z}_{{\sw}(k+1)m}.&
\end{flalign}

Therefore, if one solves Eq.~\eqref{eq:eigensystem}, 
both the angular separation constant $\Lambda_{\slmw}$ and 
the expansion coefficients $b_{jm\omega}^{(l)}$ are determined.

Once that coefficients $b_{jm\omega}^{(l)}$ are known, we can compute the spin-weighted 
spheroidal harmonics, via Eq.~\eqref{eq:spectral_decomp}, noting that the spin-weighted spherical 
harmonics can be obtained through \cite{dray}
\begin{eqnarray}
&Y_{{\sw}jm}(\chi)=\left[\frac{(2j+1)}{4\pi}\frac{(j+m)!(j-m)!}{
(j-{\sw})!(j+{\sw})!}\right]^{1/2}\left[\sin\left(\frac{\chi}{2}
\right)\right]^{2j}\non
&\times\sum_{p=p_{min}}^{p_{max}}\left(-1\right)^{j-{\sw}-p}\f{(j+{\sw})!}{p!(j+{\sw}-p)!}\non&\times\f{(
j+{\sw})!}{(p-m+{\sw})!(j-p+m)!}\left[\cot\left(\frac{\chi}{2}\right)\right]^{2p-m+{\sw}},\label{eq:spherical harmonics-1}
\end{eqnarray}
where $p_{min}=\text{max}(0,\, m-{\sw})$ and $p_{max}=min(j-{\sw},\, j+m)$. An alternative way to compute the spin-weighted spherical harmonics can be found in Refs.~\cite{hughes,Dolan:2008kf}. 

\bibliographystyle{apsrev4-1}
\bibliography{refs}
\end{document}